\documentclass[intlimits,twoside,a4paper]{article}

\usepackage[cp1251]{inputenc}

\usepackage[eqsecnum]{cmpj3}

\issue{2026}{29}{3}{33704}
\doinumber{10.5488/CMP.29.33704}
\title[Electronic energy structure and optical properties of In\textsubscript{4}CdI\textsubscript{6}]%
{Electronic energy structure and optical properties of In$_{4}$CdI$_{6}$ from $\textit{ab initio}$ calculations}
\author[I. V. Semkiv, N. T. Pokladok, F. O. Ivashchyshyn, A. I. Kashuba]{I. V. Semkiv\orcid{0000-0003-3432-8779}, N. T. Pokladok\orcid{0009-0004-3549-8205}, F. O. Ivashchyshyn\orcid{0000-0002-6919-5841}, A. I. Kashuba\orcid{0000-0003-3650-3892}\thanks{Corresponding author: \email{andrii.i.kashuba@lpnu.ua}.}}
\address{Lviv Polytechnic National University, 12 Bandera Street, 79013 Lviv, Ukraine
}

\date{Received 21 April 2026; revised 10 August 2026; accepted 11 August 2026; published 28 September 2026}

\begin{document}

\maketitle

\begin{abstract}
This work presents the $\textit{ab initio}$ calculations of the electronic energy spectrum of the In$_{4}$CdI$_{6}$ compound. The study was conducted within the framework of density functional theory (DFT), using local density approximation~(LDA) and general gradient approximation (GGA) pseudopotentials, and the Heyd--Scuseria--Ernzerhof~(HSE06) hybrid functional. The GGA approximation was implemented using the PBE and PBEsol exchange-correlation functional. Based on the electronic energy structure, the type of the minimum band gap was determined, and an analysis of the energy level dispersion for both the valence and conduction bands was performed. Furthermore, the effective masses of electrons ($\textit{m}_{c}$) and holes ($\textit{m}_{v}$) for In$_{4}$CdI$_{6}$ were established. The energy band analysis was complemented by the calculation of the density of states (DOS). Based on the electronic energy spectrum, the real and imaginary components of the dielectric function are calculated. Using the Kramers–Kronig relations, we also derive such fundamental optical functions of In$_{4}$CdI$_{6}$ as the refractive index $\textit{n}$, and the extinction coefficient~$\textit{k}$.
%
%
\printkeywords
%
\end{abstract}

\section{Introduction}


Semiconducting compounds of the A$_{4}$BX$_{6}$ group (where A = Tl, In; B = Hg, Pb, Zn, Mg, Cd, Ge; and X = Cl, Br, I) possess a wide band gap and are of significant interest due to their potential applications as materials for nonlinear optical devices \cite{1, 2, 3, 4}, ionizing radiation detectors \cite{5, 6}, temperature sensors~\cite{7}, and ion-selective electrodes \cite{8, 9}.
A$_{4}$BX$_{6}$ halides are characterised by highly pronounced anisotropies of their optical properties, ma\-king them reliable materials for optical polarizing filters, the formation of optical triggers, and related components. Their utility in nonlinear optical devices is attributed to their broad transparency windows, spanning from the visible to the far-infrared region, and to the simultaneous presence of two heavy cations within the crystal lattice, which suppresses undesirable high-energy phonon excitation.

Regarding the investigation of the crystal structure of A$_{4}$BX$_{6}$-group semiconducting compounds, such studies are well represented in the literature. Research on In$_{4}$CdI$_{6}$ originated with reference~\cite{10}. In reference~\cite{10}, it was established that the Tl$_{4}$HgBr$_{6}$ structural type describes the compounds Tl$_{4}$CdI$_{6}$ and In$_{4}$CdI$_{6}$. According to these data, the compounds crystalize in the centrosymmetric $P4$/$mnc$ structure, with lattice parameters ${a}= 906.0(2)$~pm, ${c}= 975.4(4)$~pm~\cite{10}.

Conversely, reference~\cite{11} reported on the temperature dependence of the electrical conductivity for In$_{4}$CdI$_{6}$. Based on these results, a structural phase transition was identified near $\sim210^\circ$C~\cite{11}. Furthermore, Raman spectroscopic studies of In$_{4}$CdI$_{6}$ were conducted in reference~\cite{11}. Specifically, Raman bands were observed at 24.7, 40.2, 65.9, 79.1, and 112.6 cm$^{-1}$.

Despite experimental investigations of other compounds within the A$_{4}$BX$_{6}$ group \cite{2, 3, 4, 5, 6, 7, 8, 9, 10, 11, 12, 13, 14, 15, 16, 17, 18, 19, 20, 21, 22}, data regarding In$_4$CdI$_6$ remain extremely limited. Specifically, no additional experimental reports on In$_{4}$CdI$_{6}$, beyond those mentioned above, were identified. Consequently, determining the fundamental electronic properties of the In$_{4}$CdI$_{6}$ compound represents a highly relevant task. Furthermore, theoretical studies on the electronic energy spectrum of In$_{4}$CdI$_{6}$ are also absent from the literature. This lack of data underscores the relevance and novelty of the research presented herein.

Investigations of the electronic energy spectrum of A$_{4}$BX$_{6}$ group compounds are conducted within the framework of density functional theory (DFT) \cite{4, 6, 22, 23, 24}. For instance, the energy band diagram of the Tl$_{4}$HgI$_{6}$ compound was calculated using the local density approximation (LDA) and the generalized gradient approximation (GGA) \cite{23}. The band gap values obtained from these calculations are 1.15~eV for the LDA functional and 1.265~eV for the GGA functional, both of which are somewhat underestimated compared to the experimental values. As shown in the figure, the minimum gap between the conduction and valence bands occurs at the $\Gamma$ point, the centre of the Brillouin zone (BZ), indicating that the Tl$_{4}$HgI$_{6}$ crystal is a direct-band-gap semiconductor.

The energy characteristics of Tl$_{4}$CdI$_{6}$, compared with those of Tl$_{4}$HgI$_{6}$, have been investigated in references \cite{6, 14, 20, 23}. In all cases, calculations were performed using LDA and GGA exchange-correlation potentials. The band gap values obtained in reference~\cite{23} are ${E}_{g}= 1.88$ eV (LDA) and ${E}_{g}= 2.04$ eV (GGA), which are comparable to the values of 2.043 eV (GGA) and 1.627 eV (LDA) reported in~\cite{4}. The authors of reference~\cite{6} achieved a slightly higher value (2.25 eV), through the combined use of the LDA approximation and spin-orbit coupling (SOC). The minimum band gap occurs at the centre of the Brillouin zone ($\Gamma$ point). Similar to Tl$_{4}$HgI$_{6}$, the band structure of Tl$_{4}$CdI$_{6}$ exhibits a direct character.

The investigation of the energy band diagrams for Tl$_{4}$HgBr$_{6}$ is presented in references~\cite{12, 24}. The compound under study is characterized by a direct band gap, with calculated values of 1.779~eV~(GGA) and 1.342~eV~(LDA)~\cite{12}. Additionally, a value of 2.456 eV was obtained using the mBJ+$\textit{U}$+SO approximation~\cite{12}. It is noteworthy that the latter result is in close agreement with the experimental value determined in \cite{13}.

Considering that the LDA and GGA pseudopotentials are typically used for the investigation of other compounds within the A$_{4}$BX$_{6}$ group \cite{4, 6, 22, 23, 24}, given the absence of experimentally determined energy parameters for In$_{4}$CdI$_{6}$ in the literature, this work utilises the LDA \cite{27} and GGA approximations. The GGA approximation was implemented using the Perdew--Burke--Ernzerhof (PBE \cite{25}) and PBEsol~\cite{26} parameterisations of the exchange-correlation functional. Given that hybrid functionals provide a more accurate description of the electronic energy structure of semiconducting materials, the Heyd--Scuseria--Ernzerhof (HSE06 \cite{28, 29}) functional was also employed in this study.

\section{Calculation details}

All our calculations, including geometry optimization and calculations of total energy, energy spectra and density of states, were performed with the DFT \cite{R1, R2, R3} and implemented in the QUANTUM-ESPRESSO package \cite{QE}. To describe the exchange-correlation energy of the electronic subsystem, we used the functionals taken in the LDA \cite{27}, GGA within PBE \cite{25} and PBEsol \cite{26} parameterization, and HSE06 hybrid functional \cite{28, 29}. The interaction between ions and valence electrons was described using Vanderbilt ultrasoft pseudopotentials \cite{30} for LDA and GGA. A norm-conserving pseudopotential~\cite{30} was employed in calculations performed with the HSE06 functional.

In our calculations, the value $\textit{E}_{\text{cut-off}}= 350$ eV (for LDA and GGA) and 660 eV (for HSE06) was used as the plane-wave cutoff energy (this energy corresponded to the minimum total energy). The atomic levels 4$\textit{d}^{10}$5$\textit{s}^{2}$5$\textit{p}^{1}$ for In atom, 4$\textit{d}^{10}$5$\textit{s}^{2}$ for Cd atom and 5$\textit{s}^{2}$5$\textit{p}^{5}$ for iodine atom are treated as valence electron states. The total energy converged to about $5\times10^{-6}$ eV/atom. Integration over the BZ was performed using a $2\times2\times2$ Monkhorst--Pack scheme \cite{32}. In the initial stage of our calculations, we optimized a starting In$_{4}$CdI$_{6}$ structure (with lattice parameters taken from reference~\cite{10}). The symmetry was maintained throughout the optimization process. The atomic coordinates and the unit-cell parameters were optimised using the Broyden--Fletcher--Goldfarb--Shanno technique. Optimization was continued until the forces acting on atoms were less than 0.01 eV/$\text{\AA}$, the maximum displacement was less than $5.0\times10^{-4}$ $\text{\AA}$, and the mechanical stresses in the cell were less than 0.02 GPa.

The dispersion of the energy bands was analysed using the effective masses of the electron ($\textit{m}_{c}$) and hole ($\textit{m}_{v}$), which were determined from the calculated electronic energy band structure using equation~\eqref{eq_2-1}~\cite{33}
\begin{equation}
\dfrac{1}{\textit{m}^{*}}= \dfrac{4\piup^{2}}{\textit{h}^{2}}\dfrac{\rd^{2}\textit{E}(\textbf{k})}{\rd\textbf{k}^{2}},
\label{eq_2-1}
\end{equation}
\noindent where $\textit{h}$ is the Planck constant, and $\textit{E}(\textbf{k})$ is the dependence of the band energy $\textit{E}$ on the electron wave vector $\textbf{k}$.

\section{Results and discussion}

\begin{figure}[htb]
	\center{\includegraphics[scale=0.45]{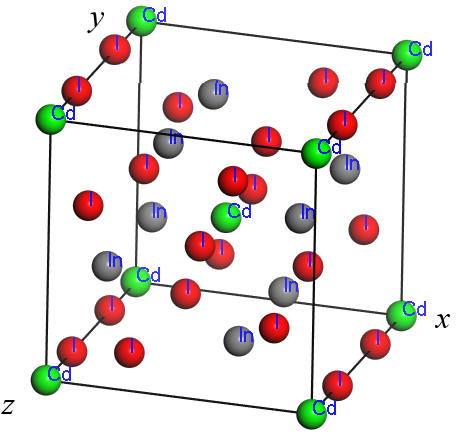}}
	\caption{(Colour online) Examples of In$_{4}$CdI$_{6}$ crystal obtained after optimization.}
	\label{f1}
\end{figure}

The optimized structures of In$_{4}$CdI$_{6}$ crystal are presented in figure~$\ref{f1}$ for visualization. It should be noted that the optimization of the In$_{4}$CdI$_{6}$ crystal by different methods leads to different lattice parameters with a tetragonal structure (space group $\textit{P}$4/$\textit{mnc}$). The space symmetry groups and the lattice parameters for the optimized structures are summarised in table $\ref{t1}$. Moreover, table $\ref{t1}$ also lists the total energy values for the optimized configurations.

A comparison of the calculated lattice parameters with the experimental data \cite{10} (see table $\ref{t1}$) shows that the values obtained using the HSE06 functional are the closest to the experimental results. Furthermore, the HSE06 hybrid functional yields the  minimum total energy ($\textit{E}_{\rm total}$) of the system.

\begin{table}[htb]
\caption{Structural properties of In$_{4}$CdI$_{6}$ crystal after optimization.}
\label{t1}
\begin{center}
\renewcommand{\arraystretch}{0}
\begin{small}
\begin{tabular}{|c|c|c|c|c|c|}
\hline
Parameters & HSE06 & LDA & GGA+PBE & GGA+PBEsol & \cite{10} \strut\\
\hline
Space group	& \multicolumn{5}{|c|}{$\textit{P}$4/$\textit{mnc}$}\strut\\
\hline
$\textit{a}= \textit{b}$, $\text{\AA}$	& 9.123182 & 10.468368	& 11.360644	& 10.471746	& 9.060(2)\strut\\
\hline
$\textit{c}$, $\text{\AA}$	& 10.160833 & 11.597852	& 12.533962	& 11.806590	& 9.754(4)\strut\\
\hline
$\textit{E}_{\rm total}$, eV	& $-18823.05$ & $-18168.19$	& $-18124.32$	& $-18088.82$	& $-$\strut\\
\hline
\end{tabular}
\end{small}
\renewcommand{\arraystretch}{1}
\end{center}
\end{table}

\begin{figure}[h]
\begin{minipage}[h]{0.49\linewidth}
\center{\includegraphics[width=1\linewidth]{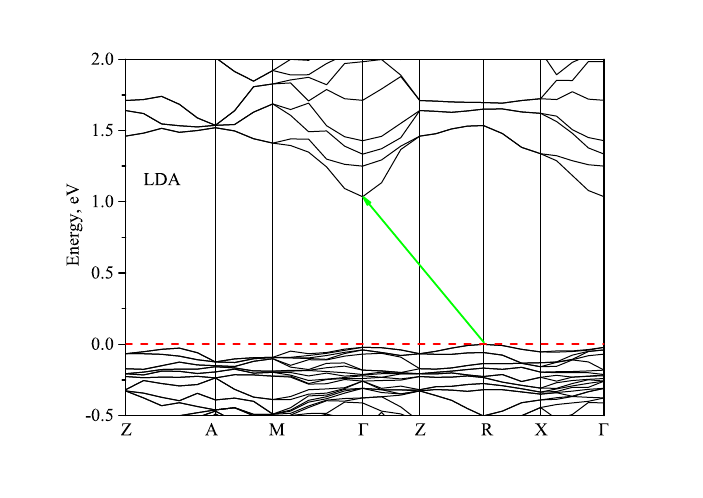}}
\end{minipage}
\hfill
\begin{minipage}[h]{0.49\linewidth}
\center{\includegraphics[width=1\linewidth]{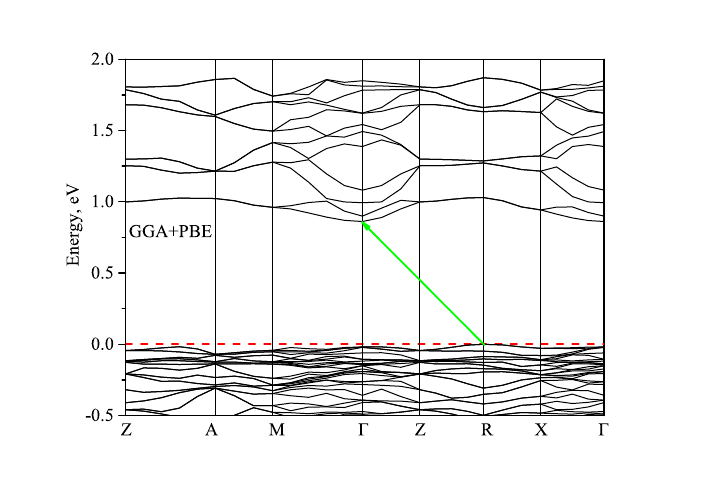}}
\end{minipage}
\hfill
\begin{minipage}[h]{0.49\linewidth}
\center{\includegraphics[width=1\linewidth]{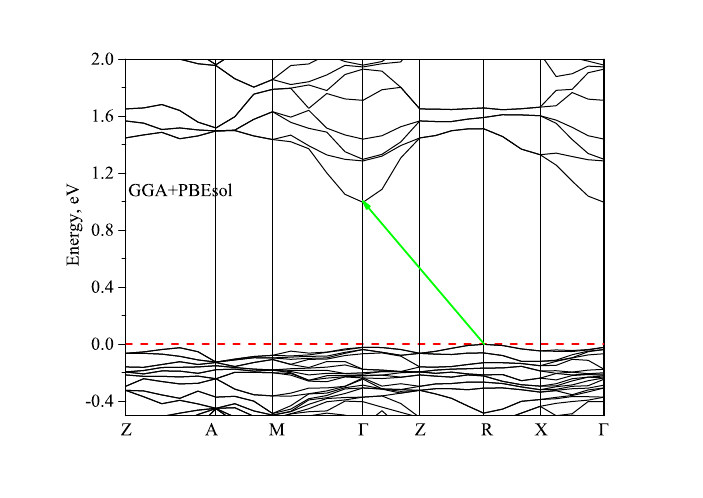}}
\end{minipage}
\hfill
\begin{minipage}[h]{0.49\linewidth}
\center{\includegraphics[width=1\linewidth]{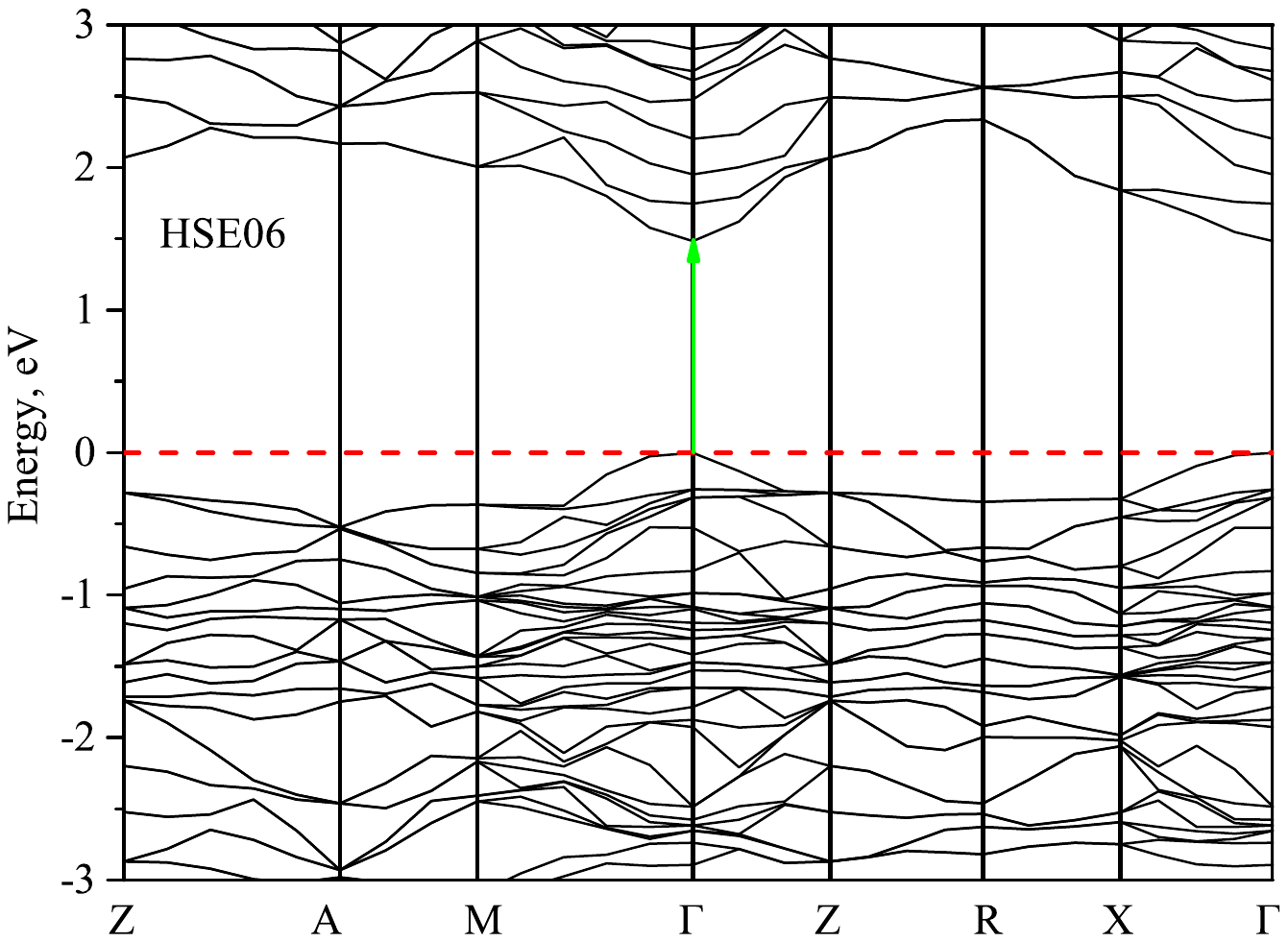}}
\end{minipage}
\caption{(Colour online) Electron band energy structures of In$_{4}$CdI$_{6}$ crystal calculated by different methods (see the legend): red lines correspond to Fermi level positions; green line --- path of the smaller band gap.}
\label{f2}
\end{figure}

Figure~\ref{f2} shows the calculated electronic energy band diagrams for In$_{4}$CdI$_{6}$ crystal along the high-symmetry points of the BZ. In this case, the energy in figure~\ref{f2} is counted from the Fermi level ($\textit{E}_{\rm F}$ = 0 eV). No experimental or theoretical data on the electronic energy spectrum of the In$_{4}$CdI$_{6}$ crystal have been identified to date. The analysis of our calculation data for the energy band spectrum of In$_{4}$CdI$_{6}$ crystal shows that the bottom of the conductivity band is localized in the centre of the BZ for all methods (HSE06, LDA, GGA+PBE, and GGA+PBEsol), i.e., at the point $\Gamma$. On the other hand, the top of the valence band is localized at the $\textit{R}$ point for LDA, GGA+PBE, and GGA+PBEsol and $\Gamma$ point for HSE06 of the BZ (see figure~\ref{f2}). This means that the corresponding crystal is characterised by an indirect energy optical band gap ($\textit{R}$--$\Gamma$) based on LDA and GGA calculations, and a direct energy optical band gap ($\Gamma$--$\Gamma$) based on HSE06. Nonetheless, the quantitative difference between the indirect optical transition and the direct transition ($\bigtriangleup\textit{E} = \textit{E}_{g}^{\rm indir} - \textit{E}_{g}^{\rm dir}$) localized at the $\Gamma$-point remains less than $\sim$0.022 eV (see table \ref{t2}). Typically, crystals of the A$_{4}$BX$_{6}$ group are characterized by a direct band gap (Tl$_{4}$CdI$_{6}$ \cite{6, 20, 23}, Tl$_{4}$PbI$_{6}$ \cite{14}, Tl$_{4}$HgI$_{6}$ \cite{23}, Tl$_{4}$HgBr$_{6}$ \cite{12, 24}). It should be noted that for the Tl$_{4}$PbI$_{6}$ crystal it shows an indirect band gap \cite{14}. This was identified using the GGA and the modified Becke-Johnson potential as implemented in the Tran-Blaha (TB-mBJ) method. However, the inclusion of the Hubbard correction parameter $\textit{U}$ and spin-orbit coupling in the calculations yielded the minimum band gap of a direct type \cite{14}. Consequently, we assumed that the direct optical band gap for  In$_{4}$CdI$_{6}$ is more correct. The experimental measurements of the optical absorption spectra for In$_{4}$CdI$_{6}$ crystal need to prove this assumption. The application of the Hubbard correction parameter $\textit{U}$ and SOC in the calculations presented in this work is considered impractical without supporting the experimental data. In the next section, we focus on the data obtained using the HSE06 method.

Regarding the minimum band gap, the value obtained with the HSE06 method was over  30$\%$ larger than those determined by the LDA and GGA methods (see table \ref{t2}). Typically, calculations performed within the LDA, GGA+PBE, and GGA+PBEsol approximations show that these approximations underestimate the band gap \cite{34}. One of the appropriate reasons for the discrepancy between theory and experiment is that infrared absorption has not been taken into account in our calculations \cite{35}. The simplest way to obtain theoretical data that are fully consistent with the experiment is to apply a so-called ‘scissors’ operator, which shifts the calculated conduction band towards higher energies, thereby increasing the band gap \cite{35}. Then, the calculated conduction band is shifted until the experimental minimum energy gap $\textit{E}_{g}$ for the compound under test is achieved. A success of the `scissors'-operator approach stems from the known Kohn--Sham equations for the band energy dispersion \cite{27}. It should be noted that the absence of an experimentally determined band gap value for this crystal precludes the performance of this operation. Given that A$_{4}$BX$_{6}$ group compounds are typically synthesized from binary precursors (which, in the case of In$_{4}$CdI$_{6}$ compound, are InI and CdI$_{2}$), their minimum bandgap is expected to lie within a range close to those of these binary constituents. For the In$_{4}$CdI$_{6}$ compound, these values are 2.01 eV \cite{36} for InI and 3.05 eV \cite{37} for CdI$_{2}$. Therefore, we assume that the bandgap value obtained using the HSE06 method is closer to the expected experimental value.

\begin{table}[htb]
	\caption{Energy properties of In$_{4}$CdI$_{6}$ crystal: the indices `$\textit{c}$' and `$\textit{v}$' denote the conduction and valence bands, respectively. ${M- \Gamma - Z}$ and ${Z-R-X}$ path of the $\textbf{k}$-space for the calculated effective masses.}
	\label{t2}
	\vspace{1ex}
	\begin{center}
		\renewcommand{\arraystretch}{0}
		\begin{footnotesize}
			\begin{tabular}{|c|c|c|c|c|c|c|c|c|}
				\hline
				\rule[-1.5ex]{0pt}{4.5ex}Method	& $\textit{E}_{g}^{\rm indir}$, eV	& $\textit{E}_{g}^{\rm dir}$, eV	& $\bigtriangleup\textit{E}$, eV & $\textit{E}_{c}^{\Gamma}$, eV	& $\textit{E}_{v}^{\Gamma}$, eV	& $\textit{E}_{v}^{R}$, eV	& $\textit{m}_{c}^{M- \Gamma - Z}$/$\textit{m}_{e}$	& $\textit{m}_{v}^{Z-R-X}$/$\textit{m}_{e}$ \strut\\
				\hline
				LDA	&1.036	&1.058	&0.022	&1.036	&$-0.02$2	&0	&2.536	&0.331\strut\\
				\hline
				GGA+PBE	&0.862	&0.881	&0.019	&0.862	&$-0.019$	&0	&3.903	&1.516\strut\\
				\hline
				GGA+PBEsol	&0.996	&1.017	&0.021	&0.996	&$-0.021$	&0	&2.594	&0.343\strut\\
				\hline
				HSE06 & 1.827 & 1.483 & $-0.344$	& 1.483	& 0	& $-0.344$ & 	0.419 &$0.233^{M- \Gamma - Z}$\strut\\
				\hline
			\end{tabular}
		\end{footnotesize}
		\renewcommand{\arraystretch}{1}
	\end{center}
\end{table}

The analysis of the energy band dispersion for the In$_{4}$CdI$_{6}$ crystal was conducted using the effective masses of electrons ($\textit{m}_{c}$) and holes ($\textit{m}_{v}$), which were determined based on equation~\eqref{eq_2-1}. Table \ref{t2} presents the effective masses for the In$_{4}$CdI$_{6}$ crystal determined from the electronic energy band diagrams along different directions in the $\textbf{k}$-space. It should be noted that the effective masses of electrons ($\textit{m}_{c}$) and effective mass of hole ($\textit{m}_{v}$) determined using the LDA and GGA+PBEsol methods are in good agreement. On the other hand, a sharp decrease in the effective masses of electrons ($\textit{m}_{c}$) and effective mass of hole ($\textit{m}_{v}$) calculated using the HSE06 method is observed compared with the LDA and GGA methods. This is attributed to the decisive impact of the HSE06 functional on the conductivity and valence band \cite{38}. Furthermore, it was found that the electron effective mass ($\textit{m}_{c}$) exceeds that of the hole ($\textit{m}_{v}$) in this crystal, consistently across different calculation methods. This disparity is expected to have a substantial impact on the sample conductivity ($\sigma\sim\textit{m}^{-1}$) \cite{33}.

\begin{figure}[h]
	\center{\includegraphics[scale=0.35]{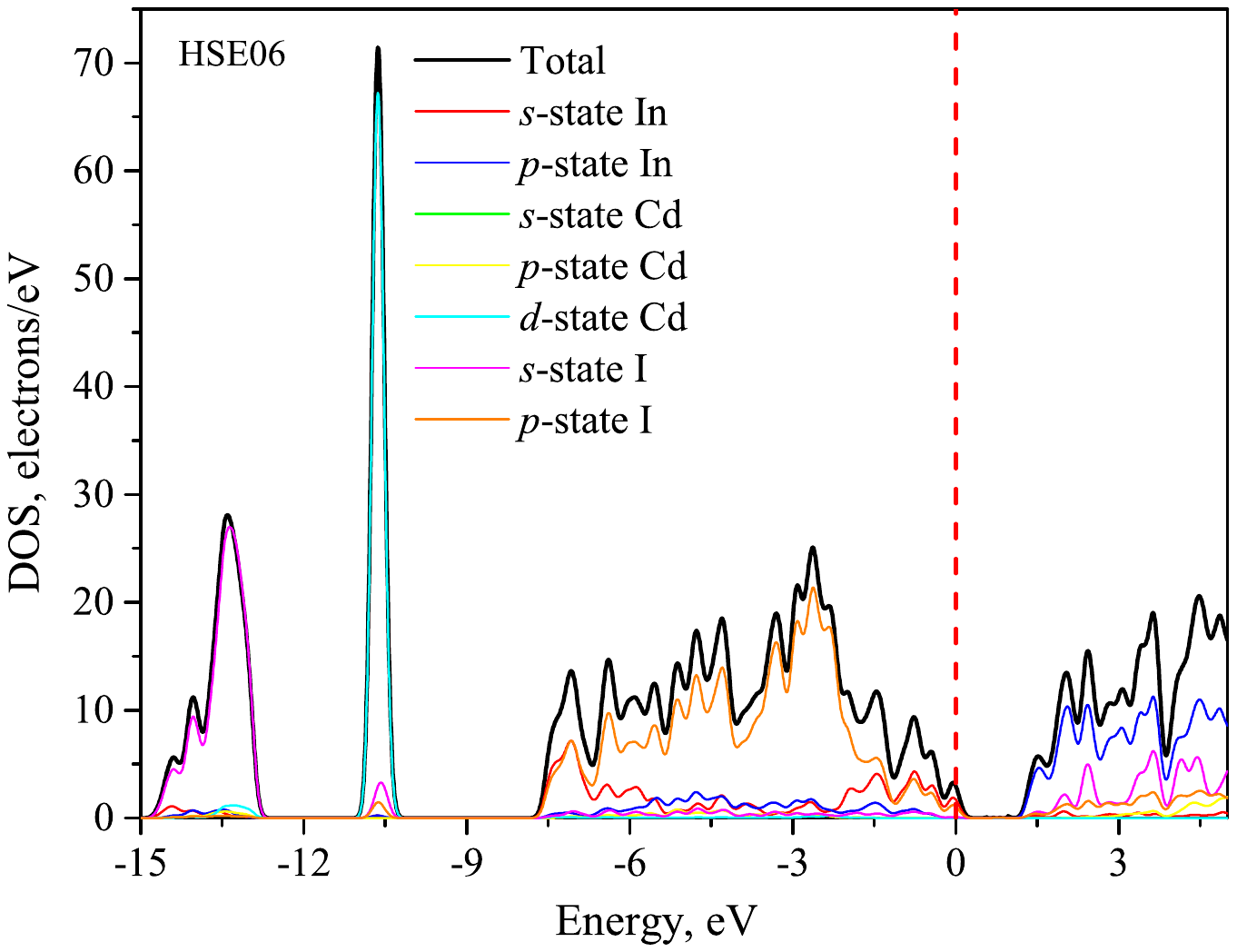}}
	\caption{(Colour online) Total and partial distributions of electronic DOS calculated for In$_{4}$CdI$_{6}$ crystal.}
	\label{f4}
\end{figure}

Based on the electronic energy structure of In$_{4}$CdI$_{6}$ crystal, we have determined the total and partial electronic density of states (DOS) (see figure~\ref{f4}). The electronic density can be summarized as follows: (i)~the energy levels between $-15$ and $-12$ eV can be attributed to the $\textit{s}$-states of iodide; (ii) the energy levels near $-11$ eV can be attributed to the $\textit{d}$-states of Cd; (iii) the $\textit{p}$-states of I are predominant in the energy region between $-8$ and $-1.5$ eV; (iv) the maximum of the valence band is formed by the contribution of the $\textit{s}$-states of In with some admixture of the $\textit{p}$-states of I; (v) the minimum of the conduction band corresponds to the $\textit{p}$-states of In; (vi) considering the Pauli exclusion principle, we infer that the minimum energy gap is formed by the $\textit{s}$--$\textit{p}$ transitions with the bond of I--In.

The total number of electrons per unit volume in the crystal can be determined from the DOS using equation~\eqref{eq_3-1} \cite{39}.
\begin{equation}
n= \int_{0}^{\infty} f(E,T,\mu)D(E) \, \rd E,
\label{eq_3-1}
\end{equation}
\noindent where $\textit{D}(\textit{E})$ is the DOS,  $\textit{f}(\textit{E}, \textit{T}, \mu)$ is the Fermi-Dirac distribution function, and $\mu$ is chemical potential. In the case of electrons:
\begin{equation}
f(E,T,\mu)=\frac{1}{1+\exp[(E-\mu)/kT]}.
\label{eq_3-2}
\end{equation}
The chemical potential is a key parameter for estimating the charge-carrier concentration ($\textit{n}$). Its limiting case, $(\textit{E}-\mu)/\textit{kT}\gg 1$, allows for the determination of the charge carrier concentration, since the Boltzmann distribution is employed as an approximation of the Fermi-Dirac distribution \cite{39}. Consequently, the charge carrier concentration can be determined using equation~\eqref{eq_3-3},
\begin{equation}
n= N_{c}\exp[-(E_{c}-\mu)/kT], \qquad N_{c}=2\left(\frac{2\piup\textit{m}^{*}\textit{kT}}{\textit{h}^{2}}\right)^{3/2}.
\label{eq_3-3}
\end{equation}
Taking into account that the $\mu = 0$ eV corresponds to the top of the valence band in semiconductors~\cite{39, 40}, the effective density of states ($\textit{N}_{c}$) and $\textit{n}$ can be estimated. Thus, the $\textit{n}$ and $\textit{N}_{c}$ were determined using the electron ($\textit{m}_{c}$) effective masses calculated in this study. The values of $\textit{N}_{c}$ and $\textit{n}$ are presented in table~\ref{t3}. It should be noted that the values of the effective density of states ($\textit{N}_{c}$) determined using methods LDA, GGA+PBE, and GGA+PBEsol are quite close to each other. By contrast, the $\textit{N}_{c}$ value estimated by the HSE06 method is different. This behaviour is attributed to the dispersion of the conduction band [$\textit{m}_{c}$, see table~\ref{t3} and equation~\eqref{eq_3-3}].

\begin{table}[htb]
	\caption{Charge carrier concentration of In$_{4}$CdI$_{6}$ crystal at 300 K: the indices `$\textit{c}$' denote the conduction bands.}
	\label{t3}
	\vspace{1ex}
	\begin{center}
		\renewcommand{\arraystretch}{0}
		\begin{small}
		\begin{tabular}{|c|c|c|c|}
			\hline
			\rule[-1.5ex]{0pt}{4ex} Method	&$\textit{m}_{c}$/$\textit{m}_{e}$	& $\textit{N}_{c}$, 10$^{20}$ cm$^{-3}$	& $\textit{n}$,  cm$^{-3}$\strut\\
			\hline
			LDA	&2.536	&1.01	& $2.25\times 10^{11}$\strut\\
			\hline
			GGA+PBE	&3.903	&1.93	&$1.22\times 10^{13}$\strut\\
			\hline
			GGA+PBEsol	&2.594	&1.04	& $5.02\times 10^{11}$\strut\\
			\hline
			HSE06	&0.419	& 0.07	& $2.53\times 10^{4}$\strut\\
			\hline
		\end{tabular}
		\end{small}
		\renewcommand{\arraystretch}{1}
	\end{center}
\end{table}

\begin{figure}[ht]
	\begin{minipage}[h]{0.49\linewidth}
		\center{\includegraphics[width=1\linewidth]{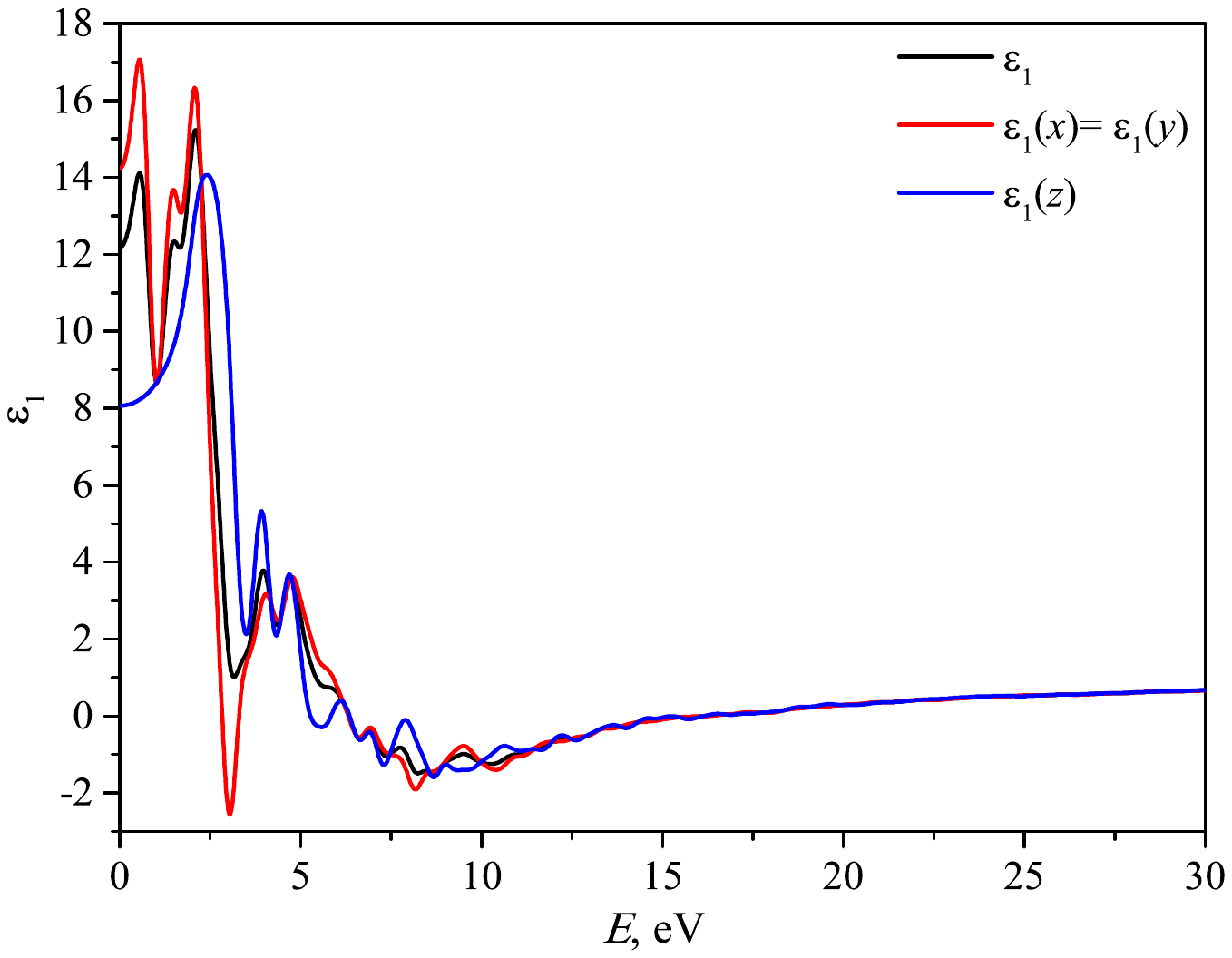}}
	\end{minipage}
	\hfill
	\begin{minipage}[h]{0.49\linewidth}
		\center{\includegraphics[width=1\linewidth]{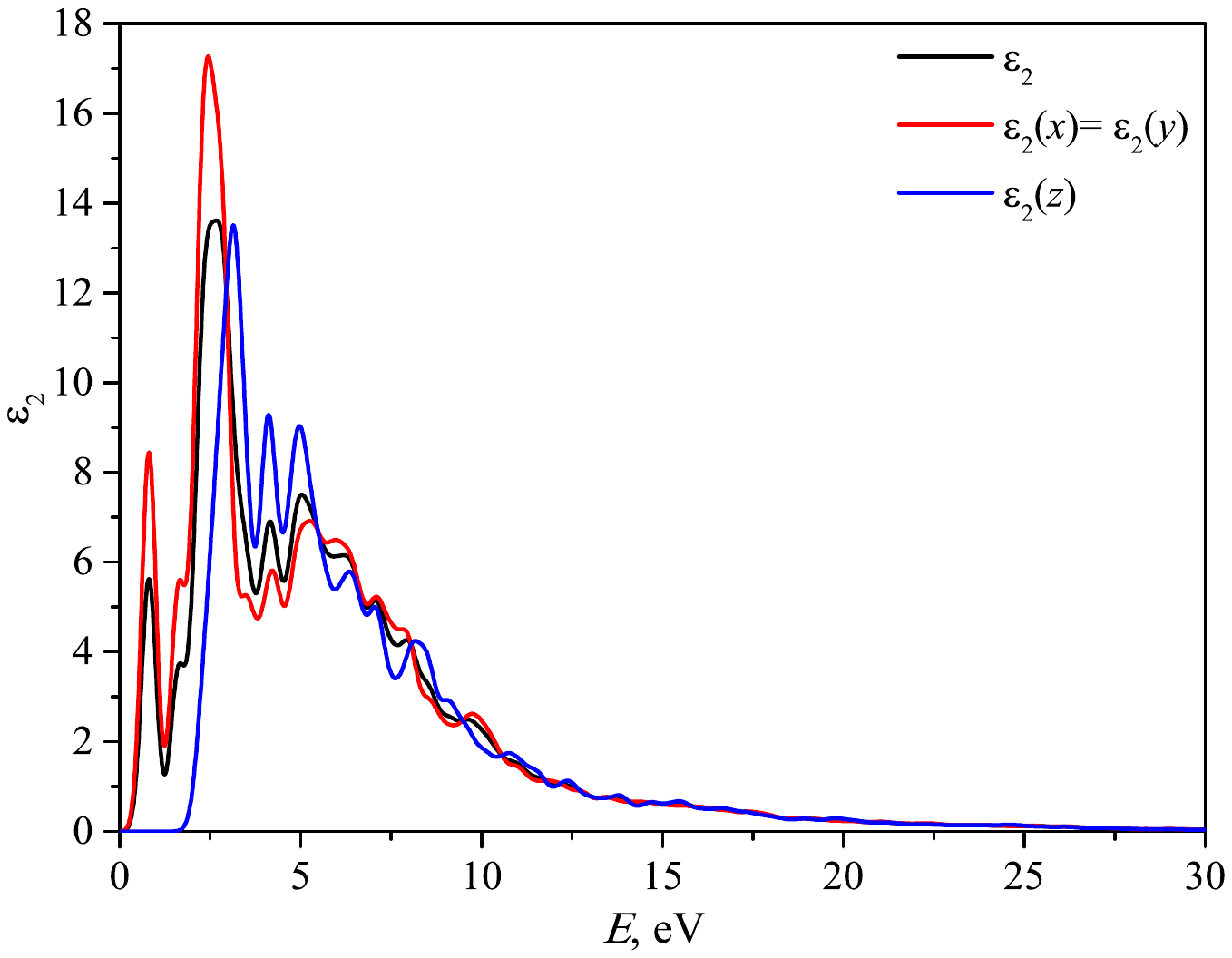}}
	\end{minipage}
	\caption{(Colour online) Real ($\varepsilon_{1}$) and imaginary ($\varepsilon_{2}$) parts of the dielectric function calculated for In$_{4}$CdI$_{6}$ crystal obtained using the HSE06.}
	\label{f5}
\end{figure}

The imaginary part of the dielectric function $\varepsilon(\hbar\omega)$ for In$_{4}$CdI$_{6}$ crystal can be calculated using the relation~\eqref{eq_3-4}~\cite{35}. Then, the real part of the function can be obtained from the Kramers--Kronig relation~\eqref{eq_3-5}~\cite{35}
\begin{align}
&\varepsilon_{2}=\frac{2e^{2}\piup}{V\varepsilon_{0}}\sum_{K,\nu,c}\vert\langle\Psi_{K}^{c}\vert\widehat{u}r\vert\Psi_{K}^{\nu}\rangle\vert^{2}\delta(E_{K}^{c}-E_{K}^{\nu}-\hbar\omega),
\label{eq_3-4}\\
&\varepsilon_{1}-1=\frac{2}{\piup}\int_{0}^{\infty}\frac{t\varepsilon_{2}(t)\rd t}{t^{2}-(\hbar\omega)^{2}}.
\label{eq_3-5}
\end{align}
In equations~\eqref{eq_3-4} and~\eqref{eq_3-4}, $\textit{E}$ is the energy, $\widehat{u}$ is the vector of polarization of the incident light, $\Psi_{K}^{c}$ and $\Psi_{K}^{\nu}$ denote the wave functions respectively of the conduction and valence bands in the $\textbf{k}$-space, $\textit{V}$ is the volume of the unit cell, $\textit{e}$ is the electron charge, $\varepsilon_{0}$ is the dielectric constant in vacuum, $\textit{r}$ is the electron position operator, $\textit{t}$ is the dummy integration variable spanning all positive photon energies, and $\hbar$ is the reduced Planck constant. The spectral dependences of $\varepsilon_{1}$ and $\varepsilon_{2}$ are presented in figure~\ref{f5}.

As shown in figure~\ref{f5}, the real part of the dielectric function increases smoothly from 0 to 2.7 eV, then decreases rapidly. At an energy near 2.9 (for $\textit{E}\perp\textit{z}$) and 5.1 eV ($\textit{E}\Vert\textit{z}$), the real part of the dielectric function turns zero and becomes negative with a further increase in energy. The imaginary part of the dielectric function is related to the photon absorption process \cite{41}. The first peak at the lowest energy corresponds to optical absorption, which is related to the  fundamental absorption of the crystal. The following bands characterize band--band transitions in a crystal at the energies higher than the band gap.

The static dielectric function is represented by a set of closely spaced overlapping bands \cite{41}. The values of the static dielectric function $\varepsilon_{0}$ are 14.29 and 8.07 for the $\textit{E}\perp\textit{z}$ and $\textit{E}\Vert\textit{z}$ directions, respectively. To numerically estimate the anisotropy of the dielectric function, we used the formula for uniaxial anisotropy \cite{41}:
\begin{equation}
\delta\varepsilon=\frac{\varepsilon_{0}^{x}-\varepsilon_{0}^{z}}{\varepsilon_{0}^{\rm tot}},
\label{eq_3-6}
\end{equation}
\noindent where $\varepsilon_{0}^{z}$ and $\varepsilon_{0}^{x}$ are the static dielectric constants and $\varepsilon_{0}^{\rm tot}$ is the total dielectric constant. The calculated value of uniaxial anisotropy of the dielectric function for In$_{4}$CdI$_{6}$ crystals is equal to $\delta\varepsilon= 0.51$.

\begin{figure}[b]
	\begin{minipage}[h]{0.49\linewidth}
		\center{\includegraphics[width=1\linewidth]{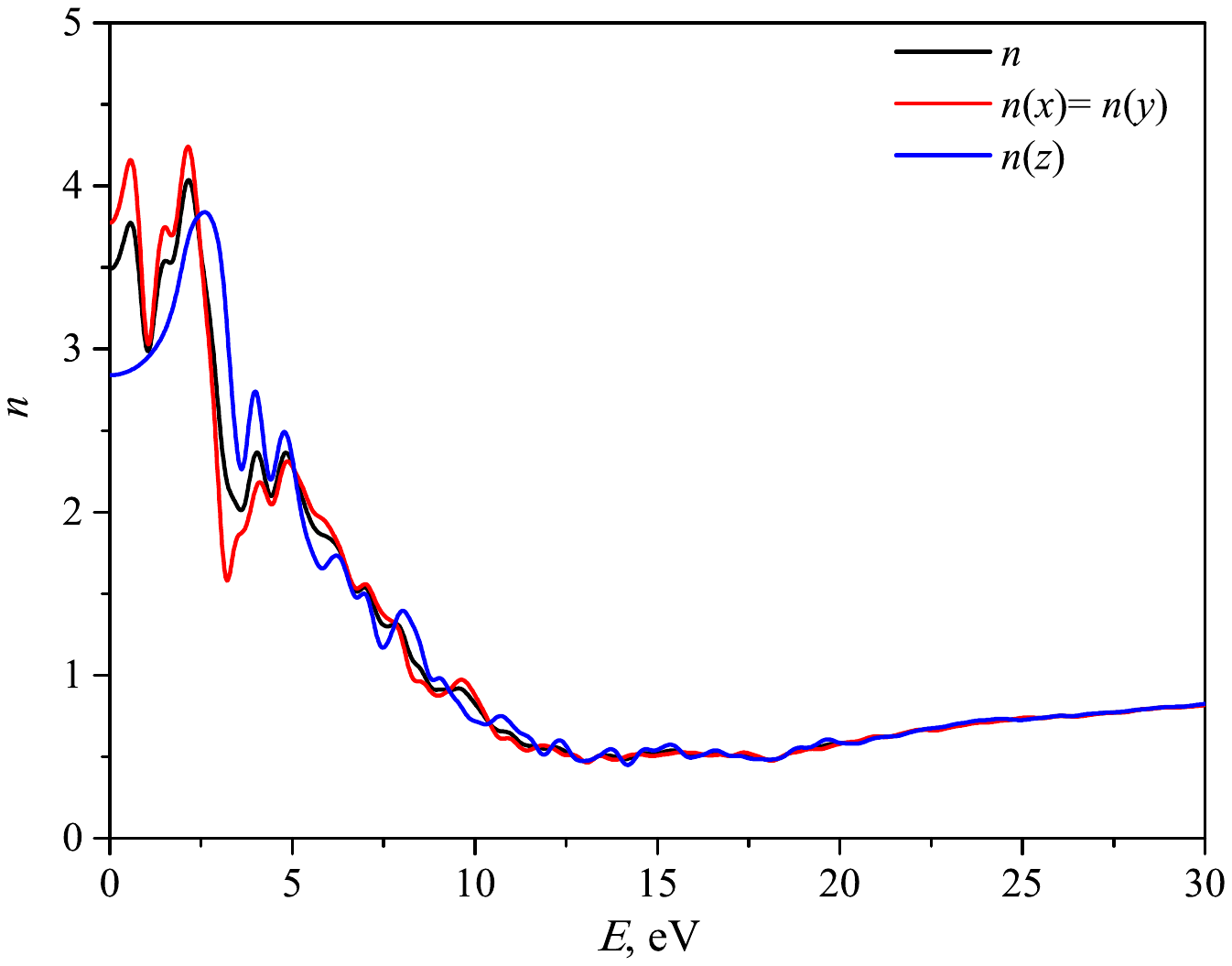}}
	\end{minipage}
	\hfill
	\begin{minipage}[h]{0.49\linewidth}
		\center{\includegraphics[width=1\linewidth]{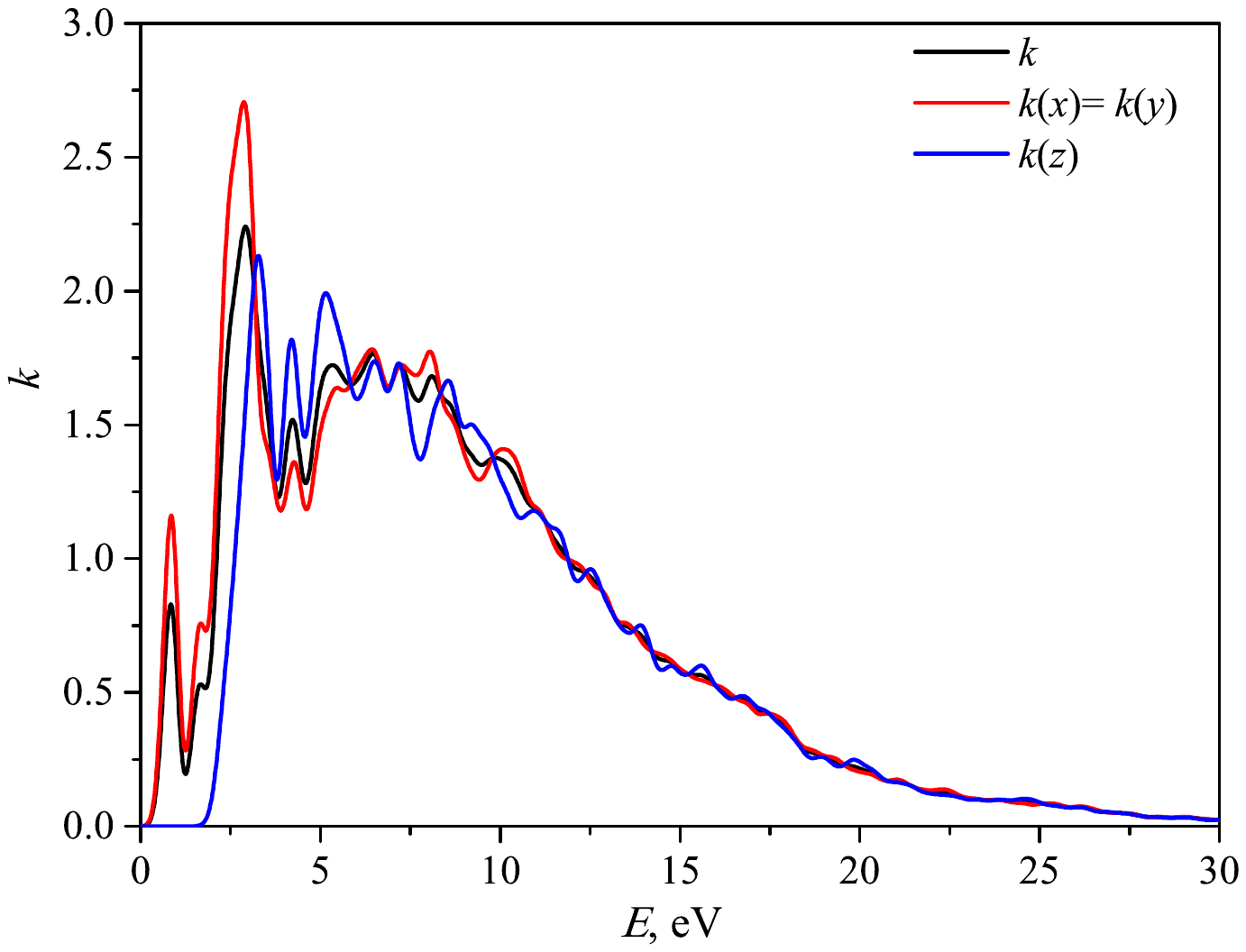}}
	\end{minipage}
	\caption{(Colour online) Calculated refractive index ($\textit{n}$) and extinction coefficient ($\textit{k}$) for the In$_{4}$CdI$_{6}$ crystal  obtained using the HSE06.}
	\label{f6}
\end{figure}

Using the calculated spectra for the real [equation~\eqref{eq_3-5}] and imaginary [equation~\eqref{eq_3-4}] parts of the dielectric function, one can obtain the following spectral dependences of the refractive index ($\textit{n}$) and extinction coefficient ($\textit{k}$):
\begin{equation}
n=\sqrt{\frac{(\varepsilon_{1}^{2}+\varepsilon_{2}^{2})^{1/2}+\varepsilon_{1}}{2}}, \qquad  k=\sqrt{\frac{(\varepsilon_{1}^{2}+\varepsilon_{2}^{2})^{1/2}-\varepsilon_{1}}{2}}.
\label{eq_3-7}
\end{equation}
Note that the spectral dependences of the refractive index and the extinction coefficient are similar to those of the real and imaginary parts of the dielectric permittivity (see figure~\ref{f6}).

\section{Conclusions}

The electronic energy spectrum of the In$_{4}$CdI$_{6}$ crystal was calculated for the first time using DFT methods (HSE06, LDA, GGA+PBE, and GGA+PBEsol). Based on the electronic energy band diagram obtained by LDA and GGA methods, it was established that the In$_{4}$CdI$_{6}$ crystal is characterized by an indirect fundamental band gap. The difference between the direct and indirect optical transitions is less than 0.022 eV. On the other hand, a direct band gap for In$_{4}$CdI$_{6}$ crystal was obtained by the HSE06 method. The dispersion of the energy bands was analysed based on the calculated effective masses of electrons ($\textit{m}_{c}$) and holes ($\textit{m}_{v}$). A significantly higher value for the electron effective mass ($\textit{m}_{c}= 0.419\textit{m}_{e}-3.903\textit{m}_{e}$) was found compared to that of the hole ($\textit{m}_{v}= 0.233\textit{m}_{e}-1.516\textit{m}_{e}$). Based on the DOS, the distribution of energy levels was determined, demonstrating that the minimum band gap is formed by $\textit{s}$--$\textit{p}$ transitions within the In–I bonds. Finally, the carrier concentration in the investigated sample was determined based on the electronic density of states. Additionally, we have calculated the spectral dependences of the primary optical functions of In$_{4}$CdI$_{6}$ crystal, i.e., the real ($\varepsilon_{1}$) and imaginary ($\varepsilon_{2}$) parts of the dielectric permittivity $\varepsilon$, the refractive index $\textit{n}$ and the extinction coefficient $\textit{k}$.

\section*{Funding}
This research was funded by Ministry of Education and Science of Ukraine, grant number 0126U000524.


\ukrainianpart

\title{Електронна енергетична структура та оптичні властивості In$_{4}$CdI$_{6}$ з розрахунків $\textit{ab initio}$}
\author{І. В. Семків, Н. Т. Покладок, Ф. О. Іващишин, А. І. Кашуба}
\address{Національний університет ``Львівська політехніка'', вул. Бандери 12, 79013 Львів, Україна}
%
%
%

\makeukrtitle

\begin{abstract}
\tolerance=3000%
У цій роботі представлені $\textit{ab initio}$ розрахунки електронного енергетичного спектра сполуки In$_{4}$CdI$_{6}$. Дослідження було проведено в рамках теорії функціоналу густини (DFT) з використанням псевдопотенціалів локального наближення густини (LDA) та загального градієнтного наближення (GGA), а також гібридного функціоналу Гейд-Скузерії-Ернцергофа (HSE06). Наближення GGA було реалізовано за допомогою обмінно-кореляційного функціоналу PBE та PBEsol. На основі електронної енергетичної структури було визначено тип мінімальної забороненої зони та проведено аналіз дисперсії енергетичних рівнів як для валентної, так і для зони провідності. Крім того, було встановлено ефективні маси електронів ($\textit{m}_{c}$) та дірок ($\textit{m}_{v}$) для In$_{4}$CdI$_{6}$. Аналіз енергетичних зон було доповнено розрахунком густини станів (DOS). На основі електронного енергетичного спектру розраховано дійсну та уявну складові діелектричної функції. Використовуючи співвідношення Крамерса-Кроніга, для кристалу In$_{4}$CdI$_{6}$ було розраховано фундаментальні оптичні функції, такі як показник заломлення $\textit{n}$ та коефіцієнт екстинкції $\textit{k}$.
\keywords A$_{4}$BX$_{6}$, електронна енергетична структура, густина станів, заборонена зона, ефективна маса.

\end{abstract}

\end{document}